\documentclass[aps,prb,showpacs,reprint,superscriptaddress]{revtex4-1}
\usepackage{graphicx}
\usepackage{amsmath}
\begin{document}


\title{Numerical adiabatic potentials of orthorhombic Jahn-Teller effects retrieved from ultrasound attenuation experiments. Application to the SrF$_2$:Cr crystal.}
 

\author{I.V. Zhevstovskikh} 
\email[zhevstovskikh@imp.uran.ru]
\affiliation{M.N. Miheev Institute of Metal Physics of Ural Branch of RAS, 620137, Ekaterinburg, Russia} 
\affiliation{Ural Federal University, 620002, Ekaterinburg, Russia}
\author{I.B.Bersuker}
\affiliation{Institute for Theoretical Chemistry, the University of Texas at Austin, TX 78712, Austin, USA}
\author{V.V.Gudkov}
\affiliation{Ural Federal University, 620002, Ekaterinburg, Russia}
\author{N.S.Averkiev}
\affiliation{A.F. Ioffe Physical Technical Institute of the RAS, 194021, St.Petersburg, Russia}
\author{M. N. Sarychev}
\affiliation{Ural Federal University, 620002, Ekaterinburg, Russia}
\author{S. Zherlitsyn}
\affiliation{Hochfeld-Magnetlabor Dresden (HLD-EMFL), Helmholtz-Zentrum Dresden-Rossendorf, D-01314 Dresden, Germany}
\author{S. Yasin}
\affiliation{Hochfeld-Magnetlabor Dresden (HLD-EMFL), Helmholtz-Zentrum Dresden-Rossendorf, D-01314 Dresden, Germany}
\affiliation {American University of the Middle East, College of Engineering and Technology, 54200 Egaila, Kuwait}
\author{G. S. Shakurov}
\affiliation{Kazan E.K. Zavoisky Physical-Technical Institute of the RAS, 420029, Kazan, Russia }
\author{V. A. Ulanov}
\affiliation{Kazan State Power Engineering University, 420066, Kazan, Russia}
\author{V. T. Surikov}
\affiliation{Institute of Solid State Chemistry of Ural Branch of the RAS, 620990, Ekaterinburg, Russia}

\date{\today}

\begin{abstract}
A methodology is worked out to retrieve the numerical values of all the main parameters of the six-dimensional adiabatic potential energy surface (APES) of a polyatomic system with a quadratic $\textit{T}$-term Jahn-Teller effect (JTE) from ultrasound experiments. The method is based on a verified assumption that ultrasound attenuation and speed encounter anomalies when the direction of propagation and polarization of its wave of strain coincides with the characteristic directions of symmetry breaking in the JTE. For the SrF$_2$:Cr crystal, employed as a basic example, we observed anomaly peaks in the temperature dependence of attenuation of ultrasound at frequencies of 50-160 MHz in the temperature interval of 40-60 K for the wave propagating along the [110] direction, for both the longitudinal and shear modes, the latter with two polarizations along the  [001] and [1$\overline{1}$0] axes, respectively. We show that these anomalies are due to the ultrasound relaxation by the system of non-interacting Cr$^{2+}$ JT centers with orthorhombic local distortions. The interpretation of the experimental findings is based on the $\textit{T}_{2g}\otimes(e_{g}+t_{2g})$  JTE problem including the linear and quadratic terms of vibronic interactions in the Hamiltonian and the same-symmetry modes reduced to one interaction mode. Combining the experimental results with a theoretical analysis we show that on the complicated six-dimensional APES of this system with three tetragonal, four trigonal, and six orthorhombic extrema points, the latter are global minima, while the former are saddle points, and we estimate numerically all the main parameters of this surface, including the linear and quadratic vibronic coupling constants, the primary force constants, the coordinates of all the extrema points and their energies, the energy barrier between the orthorhombic minima, and the tunneling splitting of the ground vibrational states. To our knowledge, such a based-on-experimental-data numerical reconstruction of the APES of a JTE problem in the five-dimensional space of all active tetragonal and trigonal displacements is realized here for the first time.
\end{abstract}


\maketitle

\section{Introduction}
Along with traditional methods of investigation of electronic structure, dynamics and local symmetry of crystal dopants and other active centers in polyatomic systems by means of experimental electron spin resonance (ESR) and pulse techniques using electron spin echo (ESE, ESEEM spectroscopy)\cite{Ref1,Ref2,Ref3,Ref4}, recent studies of II-VI:3\textit{d} and III-V:3\textit{d} crystals by means of ultrasonic experiments demonstrated the efficiency of this novel approach in obtaining important information about the structure and properties of Jahn-Teller effect (JTE) centers in such crystals \cite{Ref5,Ref6,Ref7,Ref8}. Ultrasonic experiments are uniquely suited for investigation of JTE problems because the interaction of the ultrasound wave of strain produces characteristic anomalies in speed, relaxation, and attenuation when its direction of propagation and polarization coincides with the direction of JTE distortions. Distinguished from ESR and other similar experimental studies, in which the external perturbation affects the electronic structure (indirectly influenced by the averaged nuclear dynamics), ultrasound interacts directly with the nuclear arrangement, thus allowing to reveal their adiabatic potential energy surface (APES). It was shown that ultrasound 
attenuation by JT centers with tetragonal or trigonal distortions, subject to $\textit{E}\otimes e$, $\textit{T}\otimes e $, or $\textit{T}\otimes t_2 $ problems of the JTE \cite{Ref9,Ref10}, allows for estimations of several parameters of the APES, including the JT active modes, linear vibronic coupling constants, minima positions and energy barriers between them \cite{Ref5,Ref6,Ref7,Ref8}. 
 
In these publications, rather simple cases of the JTE with either tetragonal or trigonal distortions were explored. However, in many cases JT type vibronic coupling produces much more complicated distortions. In particular, in widespread systems with threefold degenerate \textit{T} terms the JTE may lead to orthorhombic distortions involving both tetragonal and trigonal displacements \cite{Ref10}, and even lower-symmetry distortions may occur in systems with the pseudo JTE \cite{Ref10,Ref11}. A typical example of the former is the SrF$_2$:Cr crystal employed in the present paper as the basic system for ultrasonic experimentation. The Cr$^{2+}$ dopants in cubic crystals with fluorite structure (e.g., SrF$_2$, BaF$_2$, CaF$_2$, CdF$_2$) replace the ions of the bivalent metal in the lattice site of $O_h$ symmetry and are surrounded by the eight fluorine ions in the corners of the cube. The $^5D$  term of a free Cr$^{2+}$ ion is split by the cubic crystal field into the ground orbital triplet $^5T_{2g}$ and excited doublet $^5E_g$. Due to the JTE the cubic coordination of  the Cr$^{2+}$ ion in the degenerate state $^5T_{2g}$ is unstable with respect to twofold $e_g$ and threefold $t_{2g}$ type displacements of the environment \cite{Ref10}. 

If only the first coordination sphere is taken into account the problem is a two-mode one, $\textit{T}_{2g}\otimes(e_{g1}+e_{g2}+t_{2g1}+t_{2g2})$ , with a complicated APES in the space of the two twofold-degenerate $e_g$ and two threefold-degenerate $t_{2g}$ displacements that are active in this case. With the next crystalline coordination layers included the problem becomes a difficult multimode one, $\textit{T}_{2g}\otimes(e_{g1}+e_{g2}+\dots +t_{2g1}+t_{2g2}+ \dots )$, with a large number of repeating same-symmetry modes \cite{Ref10,Ref12}. Fortunately, it can be reduced to the single-mode $\textit{T}_{2g}\otimes(e_{g}+t_{2g})$ problem by applying the method of the so-called interaction mode \cite{Ref10,Ref12}, which in fact describes the real (combined) distortions produced by the same-symmetry modes resulting in the same critical points of the APES. Obviously, the parameter values of the APES retrieved from experimental data reflect the JTE distortions in terms of these summarized interaction modes. In this respect, the method described in this paper allows to reveal the solutions of the multimode problem which is very difficult to obtain by calculations only \cite{Ref10,Ref12}. 
    
In the linear approximation of the vibronic coupling the dominant interaction with one of these two types of displacements, $e_g$ or  $t_{2g}$, results in the tetragonal or trigonal minima of the APES, respectively. With the quadratic vibronic coupling terms included the JT distortions involve both trigonal and tetragonal modes producing orthorhombic minima \cite{Ref10,Ref12,Ref13}. Experimental data of ESR and ESE \cite{Ref1,Ref2,Ref3,Ref4} show that in this crystal the JTE distortions are of orthorhombic symmetry, meaning that the full $\textit{T}_{2g}\otimes(e_{g}+t_{2g})$ JTE problem with significant quadratic vibronic coupling takes place \cite{Ref13}.  With all five JT active coordinates included, two $e_g$ type and three $t_{2g}$ type, the APES becomes most complicated. The full problem is nonadiabatic, in which revealing the APES is an important first step that allows for qualitative and semi-quantitative estimates of a series of observable properties.  
  
   In the present paper we report the development of a method of reconstruction of the JTE APES in the six-dimensional space with orthorhombic global minima by retrieving its main parameters from experimental data on ultrasound attenuation, using the SrF$_2$:Cr crystal as a working example. The measurements were carried out at 50-160 MHz on crystal samples with chromium impurities of low concentration (n$_{Cr}$=1.6$\cdot10^{19}$ cm$^{-3}$). Large peaks in the temperature dependence of attenuation were found around 50 K for the waves propagating along the [110] axis for the shear mode with the polarization along the [001] (along the elastic modulus $c_{44}$) and the longitudinal mode (along the modulus $(c_{11}+c_{12}+2c_{44})/2$). A small anomaly was also observed in attenuation of the shear mode polarized along [1$\overline{1}$0]  (modulus $(c_{11}-c_{12})/2$) in the same region. We interpreted these data as due to relaxation in the system of non-interacting Cr$^{2+}$ centers, subject to the multimode JTE reduced to the  $\textit{T}_{2g}\otimes(e_{g}+t_{2g})$ problem, where $e_g$ and $t_{2g}$ are the interaction modes \cite{Ref10,Ref12,Ref13}. In combination with a theoretical analysis, we show that the global  minima of the APES have orthorhombic symmetry, and we estimate the positions and coordinates of trigonal and tetragonal saddle points, as well as the values of the linear trigonal, tetragonal, and quadratic vibronic coupling constants, the energy barrier between the orthorhombic minima, and the frequency of over-the barrier free rotations, the tunneling splitting of the ground vibrational states, and the values of the primary force constants. Such a reconstruction of the six-dimensional APES for a rather complicated JTE system based on experimental data is realized here for the first time. The developed method can be directly applied to other cases of the JTE with the five-dimensional $\textit{T}\otimes(e+t_2)$ problem, and the results obtained for the fluorite crystals are important for the search and investigation of novel materials for optoelectronics. 
  
   \section{Experimental methods and results}
The SrF$_2$:Cr single crystals were grown by Czochralski method in a helium atmosphere with small additions of thermal decomposition products of Teflon. The chromium impurity was introduced into the melt as a CrF$_3$ well-dried powder. Addition of fluorine to the atmosphere of crystal growth was aimed at creating a non-stoichiometric melt with excess of fluorine. This procedure has contributed to the increase of equilibrium concentration of chromium ions dissolved in the melt. It turned out that the presence of excess melt of fluorine ions is a prerequisite to introducing chromium in the growing crystal lattice. Depending on the crystal growth conditions, either bivalent chromium centers or centers of trivalent chromium could be formed advantageously, the charge of the latter being compensated by interstitial F$^-$ ions in the third to fifth coordination spheres around the Cr$^{3+}$ ion. A detailed study of the crystals by means of EPR \cite{Ref1} showed that the samples SrF$_2$:Cr contained mostly centers of bivalent chromium.

The Cr concentration in the SrF$_2$:Cr crystal was determined by means of an ELAN 9000 ICP-MS quadrupole-based instrument (Perkin-Elmer SCIEX) with standard operating and acquisition parameters. Solutions of calibrating standards, samples, and blanks were introduced pneumatically into ICP by GemTip cross-flow nebulizer equipped with Scott-type spray chamber. Isotopes $^{52}$Cr were used for monitoring the analytical signals. Chromium concentration proved to be n$_{Cr}$=1.6$\cdot10^{19}$ cm$^{-3}$.

Ultrasonic experiments were carried out with the help of setup operating as a variable frequency bridge \cite{Ref14}. The ultrasonic waves propagating along the [110] axis were generated and registered by a LiNbO$_3$ piezoelectric. The [110] direction was chosen because there are no degenerate normal modes propagating along such crystallographic axes, and all the non-vanishing elastic moduli  $c_{11}$, $c_{12}$, and $c_{44}$ can be measured. 

 Attenuation $\alpha$ of the modes were determined by the corresponding elastic modulus, which we take as a complex variable $c_\beta$: $\alpha_\beta=(1/2)  (\omega/v_\beta)$(Im$[c_\beta]$/Re$[c_\beta])$, $v_\beta$=Re$[\sqrt{c_\beta/\rho}]$, where $\rho$ρ denotes the density of the crystal, and $\omega$ is the cyclic frequency of ultrasound. Index $\beta$ denotes the type of the mode: one longitudinal ($\beta=l$) and two shear waves, namely, $\beta=T$ with polarization along [001], and $\beta=E$, [1$\overline{1}$0].  Accordingly, $c_l =(c_{11}+c_{12}+2c_{44})/2$, $c_T=c_{44}$, and $c_E=(c_{11}-c_{12})/2$,  where $c_{11}$, $c_{12}$, and $c_{44}$  are nonzero components of the elastic moduli tensor in the cubic crystal. The $c_T=c_{44}$ modulus in the expression for $c_\beta$ indicates that the wave produces trigonal distortions of the [CrF$_8]^{6-}$ center (i.e., distortions along the $\langle111\rangle$ type axis), whereas $c_E$ indicates tetragonal distortions (along $\langle100\rangle$). The expression for $c_l$ can be introduced in terms of symmetry moduli $c_E$, $c_T$, and $c_B=(c_{11}+2c_{12})/3$, where $c_B$ is the bulk modulus describing the totally symmetric distortions $c_l=c_B+c_T+c_E/3$. Thus the local JTE $e$ type mode manifests itself in the attenuation of the shear wave polarized along the axis [1$\overline{1}$0] and in the longitudinal wave, whereas the local  mode $t_2$ initiates the anomalies in the attenuation of longitudinal wave and the shear wave with polarization along [001].
 
\begin{figure}[ht]
 \includegraphics[width=0.95\columnwidth]   {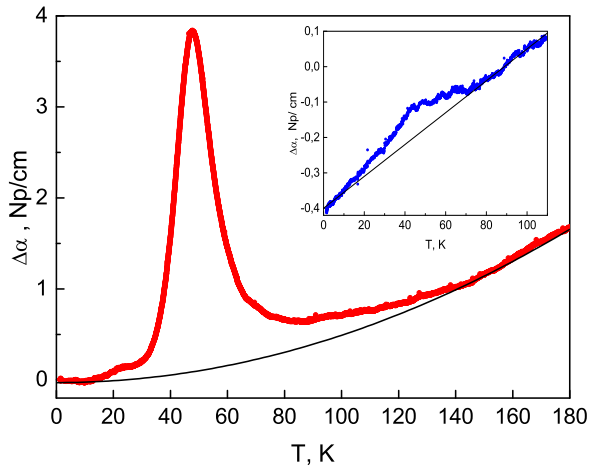}%
 \caption{Temperature dependence of attenuation of the shear ultrasonic wave at a frequency of 105 MHz  propagating along the [110] axis of the crystal SrF$_2$:Cr with the polarization along [001]. The inset shows the dependence of attenuation of the shear wave with polarization along [1$\overline{1}$0]. $\Delta\alpha(T)=\alpha(T)-\alpha(T_0)$, $T_0$=1.4 K. Solid lines are background fitting curves.}
 \label{fig1}
 \end{figure}

\begin{figure}[ht]
 \includegraphics[width=0.95\columnwidth]   {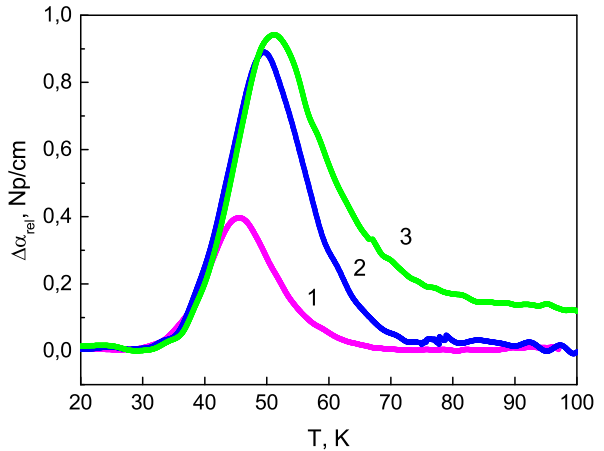}%
 \caption{Temperature dependences of the relaxation attenuation of longitudinal ultrasonic waves propagating along the [110] axis of the SrF$_2$:Cr crystal, at the wave frequencies of 61 MHz (curve 1), 142 MHz (2), 162 MHz (3). $\Delta\alpha_{rel}=\alpha(T)-\alpha_b(T_0)$ , $\Delta\alpha(T)=\alpha(T)-\alpha(T_0)$, $T_0$=1.4 K, and $\alpha_b(T)$ is the background attenuation (which is different for different frequencies).}
 \label{fig2}
 \end{figure}

The results of ultrasonic experiments are shown in Fig.\ref{fig1} and \ref{fig2}. In the temperature interval of 40-60 K, attenuation peaks are observed at different frequencies for the shear modes with polarizations along the [001] (Fig.\ref{fig1}) and [1$\overline{1}$0] (inset in Fig.\ref{fig1}) axes, as well as for the longitudinal mode (Fig.\ref{fig2}). The values of these peaks are significantly different for the waves of different polarization. The largest attenuation was found for the shear wave with the polarization along [001] which is coupled with the elastic modulus $c_{44}$, while for the wave with polarization along [1$\overline{1}$0] (modulus $(c_{11}-c_{12})/2$) the attenuation alternation is less by almost an order of magnitude. Large changes in the attenuation of the transverse mode (module $c_{44}$) contribute significantly to the attenuation of the longitudinal waves. With increase of the ultrasound frequency the position of attenuation maxima shifts to higher temperatures.    

The anomalies in the temperature dependence of ultrasound attenuation   determined by the elastic moduli $c_T$ and $c_E$ indicate the presence of trigonal and tetragonal distortion of the [CrF$_8$]$^{6-}$ center, respectively. The contribution of the totally symmetric distortions can be determined from the attenuation of the \"{}breathing\"{}  mode coupled with the bulk elastic modulus $c_B=(c_{11}+2c_{12})/3$,  expressed in terms of $c_T$, $c_E$ and $c_B=c_l-c_T-c_E/3$ {} \cite{Ref5}:
\begin{eqnarray}
\label{eq:1}
\Delta\alpha &\equiv&\frac{\omega}{2c_{l0}v_{l0}}{\rm Im}(\Delta c_B)\nonumber\\
&=&\left(\Delta\alpha_l-\frac{v^3_{T0}}{v^3_{l0}}\Delta\alpha_T-\frac{1}{3}\frac{v^3_{E0}}{v^3_{l0}}\Delta\alpha_E\right) 
\end{eqnarray}            
where the index \"{}0\"{} denotes the corresponding magnitudes at a fixed reference temperature $T_0$. Substituting in Eq.(\ref{eq:1}) the experimental data on attenuation of all the modes measured at 52 MHz we found that with an experimental error of 5\% the contribution of the \"{}breathing\"{} mode can be ignored.  
 \section{Theoretical analysis and interpretation}
\subsection{Relaxation time}
\begin{figure}[ht]
 \includegraphics[width=0.95\columnwidth]   {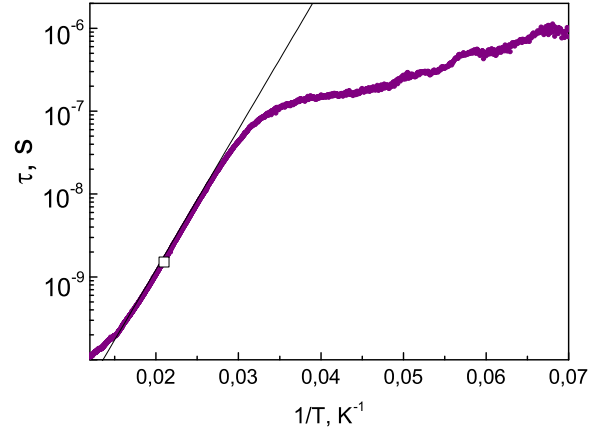}%
 \caption{Relaxation time versus inverse temperature obtained from attenuation of the shear ultrasonic wave propagating along the [110] axis with the polarization along [001] at the frequency of 105 MHz. The white square corresponds to the condition $\omega\tau=1$, and the straight line represents the function $\tau(1/T)=\tau_0\exp(V_0/kT)$ with $\tau_0=0.5\cdot10^{-12}$ sec, $V_0$ = 390 K=271 cm$^{-1}$.}
 \label{fig3}
 \end{figure}

   The method for extracting the ultrasound relaxation time $\tau(T)$   from the measured attenuation values $\alpha(T)$, based on the assumption that the peak in its temperature dependence is caused basically by the relaxation in the system of non-interacting JT centers, is explored in Ref. \cite{Ref15}. The contribution to the ultrasound attenuation by other mechanisms at low temperatures is approximated by a background monotonic function $\alpha_b(T)$ (solid lines in Fig.\ref{fig1}). Then the relaxation attenuation by the JTE centers is $\alpha_{rel}=\alpha(T)-\alpha_b(T)$,  and \cite{Ref15}
\begin{equation}
\label{eq:2}
\tau(T)=\frac{1}{\omega}\left[\frac{\alpha_{rel}(T_1)T_1}{\alpha_{rel}(T)T}\pm\sqrt{\left(\frac{\alpha_{rel}(T_1)T_1}{\alpha_{rel}(T)T}\right)^2-1}\right]
\vspace*{0.1cm}
\end{equation}
where $T_1$  is the temperature at $\omega\tau=1$, which can be determined from the position of the maximum of the function $f(T)=\alpha_{rel}(T)\cdot T$ . The function $\tau(1/T)$  obtained with the data from Eq.(\ref{eq:2}) for the shear ultrasonic wave coupled with the modulus $c_T=c_{44}$ is shown in Fig.\ref{fig3}.
 The temperature dependence of the relaxation time $\tau(1/T)$ in the interval of $T$=33-66 \textmd{K} ($1/T$=0.015-0.03 $K^{-1}$) is approximated by a straight line, which makes it possible to determine the thermal activation energy $V_0$= 390 K=271 cm$^{-1}$ and the frequency of over-the-barrier rotations $\tau^{-1}_0=2\cdot 10^{12}$ sec$^{-1}$.
\subsection{Linear vibronic coupling constants}
As noted above, the ground electronic term of the Cr$^{2+}$ ion in the cubic coordination of the SrF$_2$ crystal is a triply degenerate $^5T_{2g}$ term. The observed in our experiments trigonal and tetragonal local distortions indicate that in this case the JTE problem is $\textit{T}_{2g}\otimes(e_{g}+t_{2g})$ \cite{Ref10}, in  which the APES depends on all the five  coordinates of the interaction modes, two tetragonal of the $e_g$ mode, $Q_\epsilon$  and $Q_\theta$, and three trigonal ones for $t_{2g}$, $Q_\xi$ , $Q_\eta$   and $Q_\zeta$. In the linear approximation the potential energy operator $\hat U$ of the vibronic Hamiltonian contains only two vibronic coupling constants, $F_E$ for tetragonal and $F_T$ for trigonal displacements, respectively \cite{Ref10}: 
\begin{equation}
\label{eq:3}
\hat U=\left\|\begin{array}{ccc}
F_E(\frac{1}{2}Q_\theta-\frac{\sqrt{3}}{2}Q_\epsilon)&-F_TQ_\zeta&-F_TQ_\eta\\
-F_TQ_\zeta&F_E(\frac{1}{2}Q_\theta+\frac{\sqrt{3}}{2}Q_\epsilon)&-F_TQ_\xi\\
-F_TQ_\eta&-F_TQ_\xi&-F_EQ_\epsilon
\end{array}\right\|
\end{equation}

The solutions $\varepsilon_k(Q)$ (\textit{k}=1,2,3) of secular equation $\|\hat{U}_{\gamma\gamma\prime}-\varepsilon_k\|=0$, $\gamma$,$\gamma\prime$=1,2,3 in combination with the initial (elastic) energy describe the three branches of the APES in the five-dimensional  space of  coordinates $Q$ {} \cite{Ref16} as follows: 
\begin{eqnarray}
\label{eq:4}
E_k(Q)=\frac{1}{2}K_E\left(Q^2_\epsilon+Q^2_\theta\right)\nonumber\\ 
+\frac{1}{2}K_T\left(Q^2_\xi+Q^2_\eta+Q^2_\zeta\right)+\varepsilon_k(Q),
\end{eqnarray}
where $K_E$ and $K_T$ are the primary force constants (the force constants without the vibronic coupling contribution). Investigation of this surface in general is difficult, but its main properties can be revealed by analyzing its cross sections along specific directions.  Dependent on the relation between the vibronic constants $F_E$ and $F_T$, this surface possesses either tetragonal or trigonal minima, determining the corresponding distortions of the cubic environment of the chromium ion \cite{Ref16}. 

By taking into account the quadratic terms of the vibronic interaction (see next section), if their contribution is sufficiently strong (which is expected in the case of the SrF$_2$:Cr crystal under consideration), tetragonal and trigonal minima become the saddle points, augmented with deeper global minima of orthorhombic symmetry adjusted to them \cite{Ref13}.   Note that near the cubic symmetry the quadratic interactions are small compared with the linear ones, so they practically do not change the slope of the surface in the direction of the initial tetragonal and trigonal distortions characterized by the linear vibronic constants  $F_E$ and $F_T$. Hence we can define these constants from the experimental data on ultrasonic attenuation using the method introduced in Refs. \cite{Ref5,Ref6}. As it was shown in these papers, relaxation attenuation of the shear ultrasonic waves is given by the following expression:
\begin{equation}
\label{eq:5}
(\alpha_{rel})_i=\frac{1}{2}\frac{na_0F^2_i}{k_BT}\frac{k_{i0}}{c_{i0}}\frac{\omega\tau}{1+\omega^2\tau^2}
\end{equation}
where $k_0=\omega/v_0$ is the wave number, $v_0=v(T_0)$ is the phase velocity of the ultrasonic wave defined at $T=T_0$, $c_{i0}$ is the shear elastic modulus (the index $i=E,T$ denotes $c_E=(c_{11}-c_{12})/2$ and $c_T =c_{44}$, respectively), $n$ is the concentration of the JT centers, $a_0$ is the distance between the nearest  Cr$^{2+}$ and F$^-$ ions, $k_B$ is the Boltzmann constant. Eq. (5) at $\omega\tau=1$ yields the  expression for the linear vibronic coupling constants:
\begin{equation}
F^2_i=\frac{4k_bT_1c_{i0}(\alpha_{rel}(T_1))_i}{na_0k_{i0}}
\label{eq:6}
\end{equation}
As a result, the following values of the linear vibronic constants were obtained by means of Eq.(\ref{eq:6}): $|{F_T}|=3.66\cdot10^{-5}$ dyn and $|{F_E}|=0.63\cdot10^{-5}$ dyn. In these calculation we employed the following constants: $n=1.6\cdot10^{19}$ cm$^{-3}$, $a_0$=2.54 \r{A}  (determined from the lattice parameter $a$=5.86 \r{A}  \cite{Ref17}), $c_{11}=12.87\cdot10^{11}$ dyn/cm$^2$, $c_{12}=4.748\cdot10^{11}$ dyn/cm$^2$, $c_{44}=3.308\cdot10^{11}$ dyn/cm$^2$ {} \cite{Ref18}. The values of the linear vibronic constants show that the trigonal distortions are about 6 times larger than the tetragonal ones.
\subsection{Extrema points on the six-dimensional APES}
As mentioned above, the orthorhombic symmetry of the Cr$^{2+}$ centers in the SrF$_2$ crystal was revealed earlier in ESR and related experimental studies \cite{Ref1,Ref3,Ref19}. Following the JTE theory, in the linear $\textit{T}_{2g}\otimes(e_{g}+t_{2g})$ problem the APES has three types of the extrema points: three  tetragonal, four  trigonal, and six  orthorhombic \cite{Ref10,Ref16}. Dependent on the relation between the vibronic coupling constants and the force constants, the minima of the APES are at either tetragonal or trigonal extrema points, while the orthorhombic points are saddle points in this approximation. Sufficiently large contributions of the quadratic terms shift the APES towards orthorhombic distortions (combinations of one tetragonal and one trigonal normal displacements), transforming the six orthorhombic saddle points into global minima, and leaving the tetragonal and trigonal extrema as saddle points \cite{Ref13} (see also in \cite{Ref10,Ref12}). As mentioned above,near the cubic configuration quadratic displacements are very small as compared with the linear ones, so the linear vibronic coupling constants are not affected by quadratic terms, and the same is true for the positions of the tetragonal and trigonal extrema points produced by the linear terms (the descend toward orthorhombic minima from the trigonal extrema points goes along another cross-section of the APES). This allows us to use the following formulas for the JT stabilization energies \cite{Ref10}: 
\begin{equation}
E^E_{JT}=\frac{F^2_E}{2K_E}
\label{eq:7}
\end{equation}
for the tetragonal extrema, and
\begin{equation}
E^T_{JT}=\frac{2F^2_T}{3K_T}
\label{eq:8}
\end{equation}
for the trigonal ones. The positions and depth of the orthorhombic extrema points are strongly modified by the quadratic terms. With the dimensionless parameters \cite{Ref13}
\begin{equation}
A=\frac{W_{E,T}}{\sqrt{K_E K_T}}\quad ,\quad B=\frac{W_{E,T} F_E}{K_E F_T},
\label{eq:9}
\end{equation}
where $W_{E,T}$ is the constant of quadratic vibronic coupling, the stabilization energy of the orthorhombic extrema is \cite{Ref10,Ref12, Ref13}:
\begin{equation}
E^{OR}_{JT}=\frac{F^2_E(B^2+4A^2-4A^2B)}{8K_EB^2(1-A^2)}
\label{eq:10}
\end{equation}
\\
Eqs.(\ref{eq:7})-(\ref{eq:10}) show that the extrema-point positions on the APES depend on the values of five constants, namely, $F_E$, $F_T$, $K_E$, $K_T$, $W_{E,T}$, with some additional limitations below. The ultrasound experimental data allowed us to obtain the values of $F_E$, $F_T$ and the activation energy in the relaxation process $V_0$. Additional relations can be retrieved approximately  involving information from other experimental data. Below we present such estimates that seem to be quite reasonable, at least by orders of magnitude. The primary force constants $K_T$ and $K_E$ by definition are related to the local elastic properties of the [CrF$_8$] cluster with respect to the $T$ and $E$ type symmetrized dispalacements \textit{ in the absence of the JTE}, meaning under the conditions close to local elastic properties of the [SrF$_8$] cluster of the bulk crystal which are well described by the corresponding elastic moduli $c_T$ and $c_E$. Therefore, in a good approximation we can assume that $K_T/K_E=c_T/c_E=0.815$ . On the other hand, the $K_T$ value is defined by the local vibrational frequency $\hbar\omega_{\tau}$: $K_T=\omega_{\tau}^2\cdot M$, where M=6.43$\cdot 10^{-23}$ g is the reduced mass of the cluster [CrF$_8$]. The energies of trigonal type vibrations $\hbar\omega_{\tau}$ in various fluorites (except SrF$_2$) are given in the review \cite{Ref20}. Their values vary in the range of 110$\div$ 210 cm$^{-1}$. The same vibrational quant obtained by from optical \cite{Ref21} and ESR \cite{Ref3} studies experiments for the SrF$_2$:Co$^{2+}$ and SrF$_2$:Cu$^{2+}$ crystals amounts for $\sim$90 cm$^{-1}$. The energy of the shear acoustic phonons in SrF$_2$ is equals 99 cm$^{-1}$\ \cite{Ref21}. Taking into account all these experimental data as a pattern, it seems reasonable to assume that the values of the sought for trigonal vibrational frequency is around 90-110 cm$^{-1}$ (in the absence of the JTE the vibrational frequency of the Cr$^{2+}$, Co$^{2+}$, Cu$^{2+}$ ion is slightly larger than that of the Sr$^{2+}$ ion due to the smaller size of the former). The obtained in this way possible values of $\hbar\omega_{\tau}$  and related $\hbar\omega_{e}$, $K_T$, and $K_E$ ($\omega_e=\sqrt{{K_E}/{M}} $)  are listed in Table \ref{table1} together with the stabilization energies of the tetragonal and trigonal extrema points, calculated by Eqs.(\ref{eq:7}) and (\ref{eq:8}).
 \begin{table}[ht] 
 \caption{Approximate local vibrational frequencies, primary force constants, and tetragonal and trigonal extrema-point stabilization energies for the Cr$^{2+}$ center in the SrF$_2$:Cr crystal calculated for three close values of $\hbar\omega_{\tau}$ .}
\label{table1}
 \begin{ruledtabular}
 \begin{tabular}{|c|c|c|c|c|c|}
$\hbar\omega_{\tau}$, &$\hbar\omega_{e}$, &$K_T$, &$K_E$, &$E^E_{JT}$, &$E^T_{JT}$,  \\
cm$^{-1}$&cm$^{-1}$&dyn/cm&dyn/cm&cm$^{-1}$&cm$^{-1}$\\
\hline
90&100&1.86 $\cdot 10^4$&2.29$ \cdot 10^4$&4.36&241\\
\hline
100&111&2.30 $\cdot 10^4$&2.82$ \cdot 10^4$&3.53&195\\
\hline
110&122&2.78 $\cdot 10^4$&3.42$ \cdot 10^4$&2.92&161\\
 \end{tabular}
 \end{ruledtabular}
 \end{table}
 \begin{table*}[t]
  \caption{Numerical values of the constants $A$ and $B$, quadratic vibronic coupling constant $W_{E,T}$, and the symmetrized coordinates of \textit{one} of the three tetragonal $(E)$, four trigonal $(T)$, and six orthorhombic $(OR)$ extrema points of the six-dimensional APES of the $T_{2g}\otimes(e_g+t_{2g})$  JTE problem in the Cr$^{2+}$ center of the SrF$_2$:Cr crystal (the orthorhombic points are minima). The coordinates of the other equivalent points can be reproduced from these data by symmetry operations.}
\label{table2}
 \begin{ruledtabular}
 \begin{tabular}{|c|c|c|c|c|c|c|c|c|}
$\hbar\omega_{\tau}$, &$A$ &$B$ &$W_{E,T}$,$10^4$ &$E$ &$E$&$T$&$OR$&$OR$ \\
cm$^{-1}$&  & &dyn/cm&$Q^{(0)}_\theta$, \r{A}&$Q^{(0)}_\epsilon$,  \r{A}&$Q^{(0)}_\xi$, \r{A}&$Q^{(0)}_\theta$, \r{A}&$Q^{(0)}_\zeta$, \r{A}\\
\hline
90&-0.798 &-0.124 &-1.65 &-0.014 &0.024&0.131&-0.427&0.574\\
\hline
100&-0.826 &-0.128 &-2.1 &-0.011 &0.019&0.106&-0.408&0.532\\
\hline
110&-0.849 &-0.132 &-2.62 &-0.009 &0.016&0.088&-0.384&0.502\\
 \end{tabular}
 \end{ruledtabular}
 \end{table*}
  
Next, we note that the attenuation peak of the ultrasonic wave coupled with the $c_T$ elastic modulus is by almost an order of magnitude larger than for the wave coupled with the $c_E$ modulus (Fig. \ref{fig1}), which means that the distortions around the Cr$^{2+}$ center along the trigonal direction are overwhelming. The same conclusion follows from the above-noted significantly (about 6 times) larger linear trigonal vibronic coupling constant than the tetragonal one. From Table \ref{table1} we see that the stabilization energy of the trigonal, $E^T_{JT}=(161\div 241)$ cm$^{-1}$, and tetragonal, $E^E_{JT}=(2.92\div 4.36)$ cm$^{-1}$, extrema points are significantly smaller than the activation energy $V_0$=271 cm$^{-1}$ obtained from the temperature dependence of the relaxation time shown in Fig \ref{fig3}. Consequently, they are lower than the potential barrier between the minima $\delta=V_0+E_0$ , where $E_0\approx 50\div$61 cm$^{-1}$ is the zero-vibration energy in the orthorhombic minima calculated in the next section. It follows that the trigonal-type extrema points on the APES (moreover, the more shallow tetragonal points) are not global minima, which in turn means that the global minima of the APES for the Cr$^{2+}$ center in the SrF$_2$ crystal are orthorhombic. The latter descend from the trigonal saddle point toward one of the tetragonal displacements, so their stabilization energy is  $E^{OR}_{JT}=E^T_{JT}+\delta$. In this case the regions of the parameters $A$ and $B$, for which the orthorhombic extrema become absolute minima, are defined by the relations (Eqs.17 in \cite{Ref13}):
\begin{align}
B=\left[-2A^2\pm 2\sqrt{3(1-A^2)}\right]/(3-4A),B<0 \mbox{ or } B>1
\label{eq:11}
\end{align}
\begin{equation}
B=2A^2\pm\frac{2}{3}A\sqrt{3(1-A)},\quad \sqrt{3}/3-1< B< 1.
\label{eq:12}
\end{equation}            
            
Taking into account that $(A/B)^2=(3/4)E^T_{JT}/E^E_{JT}$  and presenting Eq.(\ref{eq:10}) in the form
\begin{equation}
\frac{E^{OR}_{JT}}{E^E_{JT}}=\frac{(A/B)^2-(A/B)A+1/4}{(1-A^2)}
\label{eq:13}
\end{equation}
we get a quadratic equation with respect to $A$. One of its solution with $A/B>$0 calculated with the given above $\hbar\omega_{\tau}$ values occurs in the region of $B<0$ which satisfies the condition (\ref{eq:11}) for orthorhombic minima. 

    Next, we estimate the numerical values of the coordinates of the extrema points of the five-dimensional APES. They are given by the following expressions \cite{Ref10,Ref12}. For tetragonal points $Q^{(0)}_\xi=Q^{(0)}_\eta=Q^{(0)}_\zeta=0$, $Q^{(0)}_\theta=-F_E/2K_E$ and $Q^{(0)}_\epsilon=\sqrt{3}/2K_E$, while for trigonal points $Q^{(0)}_\theta=Q^{(0)}_\epsilon=0$ and $Q^{(0)}_\xi=\pm Q^{(0)}_\eta=\pm Q^{(0)}_\zeta=2F_T/3K_T $; for orthorhombic points $Q^{(0)}_\xi=Q^{(0)}_\eta=Q^{(0)}_\epsilon=0 $, $ Q^{(0)}_\theta=-F_E(B-2A^2)/2K_EB(1-A^2)$ and $Q^{(0)}_\zeta=F_T(2-B)/2K_T(1-A^2) $. With the numerical values above, and assuming that $F_T>0$, $F_E>0$, we got the numerical values of these coordinates listed in Table \ref{table2} (note that coordinates $Q$ are linear combinations of the Cartesian coordinates of atomic displacements).

\subsection{Energy barrier and tunneling splitting}
As mentioned above, the linear dependence of the relaxation time on invers temperature in Fig.\ref{fig3} indicates that the relaxation mechanism in this temperature interval is best described by thermal over-the-barrier transitions following equation $\tau(1/T)=\tau_0\exp (V_0/kT)$ , where $V_0$ = 390K = 271cm$^{-1}$ is the relaxation activation energy and $\tau^{-1}_0=2\cdot 10^{12}$  sec$^{-1}$ is the frequency of the over-the-barrier \"{} free \"{} rotations between the equivalent minima. The full barrier height employed in the previous section, $\delta=E_0 + V_0$, includes also the energy $E_0$ of the corresponding zero vibrations in the orthorhombic minimum. It can be estimated from the same full solutions of the quadratic $\textit{T}_{2g}\otimes(e_{g}+t_{2g})$ problem \cite{Ref12,Ref13}. As mentioned above, the orthorhombic minima are formed by a combination of one trigonal $t_{2g}$ displacement (e.g., along the diagonal of the cube) leading to one of the four trigonal saddle point, continued by one tetragonal displacement of $e_g$ type (e.g., along one of the contiguous lateral sides) resulting in one of the equivalent orthorhombic minima with one $e_g$ type $Q^{(0)}_\theta$ and one  $t_{2g}$ type $Q^{(0)}_\zeta$ displacements (Table \ref{table2}). Starting with the state in the minimum, to overcome the barrier to another minimum the system should go back to the trigonal saddle point via the same $e_g$ type vibrations, which are of $b_1$ symmetry, split from the cubic $e_g$ type and modified by the quadratic vibronic coupling terms. The theory \cite{Ref12} yields the following expression for the modified vibrational frequency $\omega_{B1}$:
\begin{equation}
\omega_{B1}^2=\omega^2_E\left[1-\frac{3B^2{(1-A^2)}^2}{A^2{(2-B)}^2}\right],
\label{eq:14}
\end{equation}
where $\omega_E$ is given in Table \ref{table1}. With the parameter values in Table \ref{table2} we get the value of $ \hbar\omega_{B1}$, the zero vibration energy $E_0=(1/2)\hbar\omega_{B1}$, and the total energy barrier between the orthorhombic minima $\delta=V_0+E_0$ (Table \ref{table3}).

	With the data obtained in this paper for the Cr$^{2+}$ centers in the SrF$_2$:Cr crystal we are able to predict the splitting $\delta_0$ of the ground state vibrational energy levels in the orthorhombic minima due to the tunneling between them, at least by order of magnitude. The general theory of the tunneling splitting in JT systems was first worked out in \cite{Ref22}, and for the tunneling between the orthorhombic minima of the full $\textit{T}_{2g}\otimes(e_{g}+t_{2g})$  problem under consideration the following expression for the tunneling splitting $\delta_0$ of the ground vibrational state was obtained under the approximation $K_T\approx K_E$ \cite{Ref12,Ref13}:
\begin{align}
\delta_0& =\frac{3}{2}\delta\frac{24-28B-2B^2+9B^3}{(4-3B)(4-3B^2)}\frac{S}{1-4S^2} , \nonumber \\ S& =\exp\left[-\frac{3}{2}\frac{\delta}{\hbar\omega_{B1}}\frac{12-20B-11B^2}{(4-3B)(4-3B^2)}  \right].
\label{eq:15}
\end{align}
Inserting the numerical values of the constants obtained above we get the values of $\delta_0$ listed in Table \ref{table3}. Note, however, that the numbers for $\delta_0$ are rather rough approximate because of the approximations used in the deduction of Eq.(\ref{eq:15}), and the exponential dependence of $\delta_0$ on the parameters that are known approximately. But they are reasonable by orders of magnitude.  
 \begin{table}[ht] 
 \caption{The ground state vibrational frequency in the orthorhombic minima $\omega_{B1}$, zero vibration energy $E_0$, total barrier $\delta$, stabilization energy (depth) of orthorhombic minima $E^{OR}_{JT}$, and tunneling splitting $\delta_0$ in JT centers of SrF$_2$:Cr for several $\hbar\omega_{\tau}$ values.}
\label{table3}
 \begin{ruledtabular}
 \begin{tabular}{|c|c|c|c|c|c|}
$\hbar\omega_{\tau}$, &$\hbar\omega_{B1}$, &$E_0$, &$\delta$, &$E^{OR}_{JT}$, &$\delta_0$,  \\
cm$^{-1}$&cm$^{-1}$&cm$^{-1}$&cm$^{-1}$&cm$^{-1}$&cm$^{-1}$\\
\hline
90&99.8&49.9 &321&561&13.3\\
\hline
100&110.7&55.4&326&552&19.0\\
\hline
110&122&61&332&493&25.6\\
 \end{tabular}
 \end{ruledtabular}
 \end{table}
\section{Conclusions}
   Following the worked out in this paper methodology, the six-dimensional APES of a rather complicated $T$-term JTE problem with strong quadratic vibronic coupling was revealed numerically based on the experimental data on ultrasonic attenuation (numerical values were obtained for all its main parameters). The experiments were performed on a SrF$_2$:Cr$^{2+}$ crystal, taken as an example in which the Cr$^{2+}$ ions are subjects to the multimode $\textit{T}_{2g}\otimes(e_{g}+t_{2g})$ JTE problem with  orthorhombic distortions of the environment. By choosing the directions of the ultrasound wave propagation and polarization with its strain along the JTE distortions, we explored the anomalies in the temperature dependence of wave attenuation and worked out the method of extracting the APES parameters from the experimental data in combination with the theoretical findings for this problem. All the extrema points of the six-dimensional APES of its JTE, three tetragonal and four trigonal extrema points and six orthorhombic minima, their coordinates and stabilization energies, the energy barrier between the orthorhombic minima, and the tunneling splitting of the ground vibrational states, as well as the linear and quadratic vibronic coupling and primary force constants, were estimated numerically. An important point is that these numerical data reflect the solution of a multimode JTE problem which is very difficult to obtain by calculations only.
   
The methodology worked out in this paper can be applied to other systems with complicated JTE problems, while the specific data obtained for impurity fluorites are important for the search and study of novel optoelectronic materials.

\begin{acknowledgments}
The research was carried out with the state assignment of FASO of Russia (the theme “Electron” № 01201463326), supported in part by RFBR (project №15-02-02750 a) and by UrFU Center of Excellence “Radiation and Nuclear Technologies” (Competitiveness Enhancement Program). We acknowledge the support from HLD at HZDR, member of the European Magnetic Field Laboratory (EMFL).
\end{acknowledgments}


\end{document}